\documentclass[aps,prl,floats,twocolumn]{revtex4}
\usepackage{bm}
\usepackage{epsfig}
\usepackage{rotating}

\newcommand{\nn}{\nonumber}
\newcommand{\beq}{\begin{equation}}
\newcommand{\eeq}{\end{equation}}
\newcommand{\bea}{\begin{eqnarray}}
\newcommand{\eea}{\end{eqnarray}}
\newcommand{\ben}{\begin{eqnarray*}}
\newcommand{\een}{\end{eqnarray*}}

\newcommand{\lambdaqcd}{\Lambda_{\rm QCD}}

\def\D0{D\O}

\begin{document}

\title{Anomalous 3-jet and Heavy Quark Fragmentation}

\author{Yu Jia}
\affiliation{Department of Physics and Astronomy, Michigan State University, 
	East Lansing MI 48824}

\date{\today}

{\begin{abstract}

Heavy hadron production in $e^+e^-$ annihilation
is studied in the framework of heavy-quark recombination 
mechanism.  This mechanism predicts a new type of 3-jet event
such as $b\overline{b}q$ or  $b\overline{b}\overline{q}$,
which is power-suppressed relative to the standard
$b\overline{b}g$ event.
Furthermore, heavy quark fragmentation function also receives 
a contribution from this mechanism, which is 
{\it calculable}  in perturbative QCD as long as the momentum fraction $z$ 
is not too close to unity.
We conjecture that the height of the peak of 
fragmentation function for S-wave heavy hadron 
scales as $\alpha_s^2(m)\,m/\lambdaqcd$.

\end{abstract} } 

\maketitle

\vspace{0.33 in}

Most perturbative QCD studies have focused on the leading power
contribution, guided by the factorization theorems~\cite{Collins:gx}. 
Power corrections usually pose  a big challenge, either because 
they may not be factorizable, or even if so, many more 
nonperturbative matrix elements need to be involved, thus making
phenomenological analysis more difficult. 

However, there are situations where power corrections play an essential
role. Deep inelastic scattering process at low $Q^2$ provides such 
an example~\cite{Politzer:1980me}.  Another classical  example 
is the charm production asymmetries observed in various fixed-target 
experiments~\cite{Aitala:1996hf}, where one must resort to a new kind of hadronization 
mechanism other than the fragmentation. 

Inspired by the Nonrelativistic QCD factorization~\cite{Bodwin:1994jh},
a new hadronization mechanism,  dubbed {\it heavy-quark recombination} 
(HQR)~\cite{Braaten:2001bf}, 
has been developed to explain these charm asymmetries. This mechanism simply 
exploits the fact that after a hard scattering, a heavy quark may
capture a nearby
light parton which 
is produced in the hard scattering and
happens to carry soft momentum in
its rest frame,
subsequently they can
materialize into a heavy hadron, plus additional soft hadrons.
The inclusive production of heavy hadron in this mechanism can be expressed as
a product of hard-scattering parton cross section and a 
nonperturbative parameter,
which characterizes the probability for the heavy quark and light parton to combine into
a heavy hadron. A typical example for
the HQR process in hadron collisions
is $\overline{q}\,g\rightarrow \overline{B}+\bar{b}+X$ 
(We denote $b$-flavored meson by $\overline B$).
Actually, HQR is analogous to an earlier 
higher-twist model for the light hadron production~\cite{Berger:1980qg},
e.g. $q\,g\rightarrow \pi q$.
In contrast,
this model involves a 
light-cone distribution function of $\pi$ 
convoluted with the hard-scattering amplitude, 
thus more complicated than
the multiplicative recombination factors in HQR.
Roughly speaking, it is due to simpler dynamics of a heavy hadron.
In a heavy-light hadron, most of its energy is carried by the heavy quark,
and the light partons carry a characteristic momentum of order $\lambdaqcd$.
However, the momentum of a light hadron is widely distributed among its constituent partons.

The physical idea of HQR is quite general, and not only confined in the hadro-
and photoproduction of heavy hadron. 
In this Letter, we will apply the HQR to the heavy hadron production in $e^+e^-$
annihilation.
Heavy flavor production in the linear collider program is of great interest,
because it is a perfect arena to test perturbative QCD, 
and perhaps more importantly, 
because some related
precision electroweak observables like forward-backward asymmetry
of $b$ quark may be the sensitive probes 
to the new physics~\cite{Hagiwara:fs}. 
To interpret the experimental results correctly, 
a comprehensive understanding 
of hadronization of heavy quark is compulsory.

To simplify the analysis, we will focus on the bottom meson production from  $Z^0$ decay.
Extending to the heavy baryon production, and considering the $e^+e^-$ 
center-of-mass energy off the $Z^0$ peak, will be straightforward.

\begin{figure}[bt]
  \centerline{\epsfysize= 3.1 truecm \epsfbox{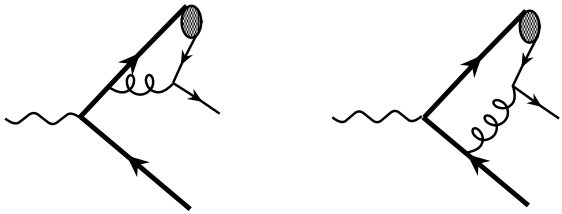}  }
 {\tighten
\caption{Diagrams for the HQR process
$ Z^0 \rightarrow b\overline{q}(n) + \overline{b}+q$. 
Wavely, thick and thin
lines represent $Z^0$ boson, bottom and light quark, respectively.
The shaded blob represents the hadronization of $b \overline q(n)$
into $\overline B$ meson plus additional hadrons.
We have suppressed two additional diagrams in which 
the light quark couples to $Z^0$,  since they don't 
contribute to the HQR cross section in the leading $1/m_b$ expansion.}
\label{zbb} }
\end{figure}

At order $\alpha_s^2$, bottom can be produced through $Z^0 \rightarrow b \overline{b} q
\overline q$.
If each quark independently fragments into hadrons, then it represents a regular 4-jet event.
Nevertheless, in a small corner of phase space while $\bar q$ is soft in 
the $b$ rest frame,
they can be projected onto an intermediate $b\overline{q}$ 
state with definite color and angular momentum. After a time of order $O(\lambdaqcd^{-1})$,
this composite state hadronizes into a $\overline B$ meson plus soft hadrons.
Therefore, we end up with a  jet containing $\overline B$  from the recombination, 
the recoiling $\bar b$ jet and a light quark jet. The corresponding Feynman diagrams are
shown in Fig.~\ref{zbb}. 
Actually they can be obtained by crossing the diagrams for the HQR process of
heavy meson photoproduction~\cite{Braaten:2001bf}.
A striking signal of this anomalous 3-jet event is that the third jet 
is initiated by a {\it light quark}, instead of by a {\it gluon}. 
The inclusive $\overline B$ production rate from HQR can be written
\bea
d \Gamma [\overline{B}] = \sum_n d{\hat\Gamma}[Z^0\rightarrow b\overline{q}(n) + \overline{b}+q]\,
\rho[b\overline{q}(n) \rightarrow \overline{B}] \,. 
\eea 
%
%where $d{\hat\Gamma}[Z^0\rightarrow b\overline{q}(n) + \overline{b}+q]$
$d{\hat\Gamma}$
is the perturbatively calculable parton cross section, 
and $\rho_n$ are the recombination factors, where $n$ denotes
the color and angular momentum of $b\overline{q}$.
We will concentrate on the S-wave states in this work.
These $\rho_n$ parameters have been recently defined 
in terms of nonperturbative matrix elements~\cite{Chang:2003ag}.
An important property of these parameters is that they scale as $\lambdaqcd/m_b$.
Since $\overline{B}$  is produced inclusively, the quantum numbers of $b\overline{q}$
don't necessarily need match those of $\overline{B}$.
However, it is found that all the existing charm meson 
asymmetries data can be fairly accommodated by a single 
color-singlet
parameter $\rho_1^c=0.15$~\cite{Braaten:2001bf}. 
Since it was fit by using only the leading order 
$gg\rightarrow c \bar c$ matrix elements as input, 
we should rescale it by a K factor to take into account 
the NLO effects.
We choose $\rho_1^c = 0.3$, which can be readily converted into
$\rho_1^b \approx \rho_1^c\,m_c/m_b = 0.1$.

The momenta of $b\overline{q}$, $\overline{b}$ and $q$ are 
assigned as $p$, $\overline{p}$ and  $q$.
It is convenient to define two independent dimensionless variables 
$y=2 \overline{p} \cdot q/M_Z^2$
and $\overline{y}=2 p \cdot q/M_Z^2$. They are related to the 
standard energy fraction variables
$z= 2p^0/M_Z$, $\bar z= 2 \overline{p}^0/M_Z$ by $y=1-z$ and $\overline{y}=1-\bar z$.
We also define  $\gamma= m_b^2 / M_Z^2$.
In the zeroth order of $1/m_b$ expansion, we can neglect the mass
differences between $\overline B$ and $b$.
Following the prescription for 
calculating the HQR contributions~\cite{Braaten:2001bf}, 
we  obtain the inclusive $\overline B$ and $\overline{B}^*$ production rate
\bea
\label{ld1s0}
{1\over \Gamma_0}{d \Gamma [\overline B]
 \over 
dy d \overline y } &=&
{64 \rho_1^b \alpha_s^2 \over 81}\gamma \left\{
 {1\over y \,\overline{y}^2}- 
 {\gamma (1+y) \over  \overline{y}^3} +
\cdots \right\},  
\\
{1\over \Gamma_0}{d \Gamma [\overline{B}^*] 
\over 
dy\,d\overline{y} } &=&
{64\rho_1^b \alpha_s^2 \over 81} \, \gamma \,\left\{
 {3-4\, y+2\,y^2\over y \, \overline{y}^2}
 \right.
\label{ld3s1}
 \\
 && \,- \,  \left.
 {\gamma \, (3-\,y) \over \overline{y}^3}
\, +\cdots \,\right\}\,, \nn
\eea
%
%where $\Gamma_0\equiv \Gamma[Z^0\rightarrow b\overline{b}]$ 
where $\Gamma_0$
is the lowest order decay rate of $Z^0$ to $b\overline{b}$.
The $b$ mass in $\Gamma_0$ is neglected for simplicity. 
The common factor $\rho_1^b$ in Eq.~(\ref{ld1s0}) and (\ref{ld3s1})
is a manifestation of heavy quark spin symmetry~\cite{Isgur:vq}.
$CP$ invariance indicates the cross sections for 
$\overline{b}q$ recombination is identical to those for 
$b\overline{q}$ recombination.  
These cross sections are highly asymmetric in $y$ and $\overline{y}$,
which is a generic feature of HQR processes.
Greater enhancement at small $\overline{y}$ relative to $y$ indicates
that light quark $q$ prefers comoving with $\overline B$ 
to with the recoiling $\bar b$. 
The ellipsis in Eq.~(\ref{ld1s0}) and (\ref{ld3s1}) represent 
terms that are less singular at small  $\overline{y}$. 
Generally, they
also  have different coefficients for the vector coupling 
and axial-vector coupling, signalling 
a nonzero quark mass breaks the chiral symmetry. 
Full expressions of these cross sections will be given 
somewhere else.

In a $b\overline{b}g$ event, perturbative calculation breaks down
when the outgoing $g$ becomes soft, which corresponds to an
almost on-shell $b$ propagator.
Similarly, when the outgoing $q$ becomes soft, 
perturbative calculation for the HQR process will also become invalid.
To ensure that the HQR parton process is governed by the
short-distance scale, besides the restriction on the virtuality of $b$ propagator, 
the virtuality of the gluon that
splits into $q \overline q$ must satisfies 
$(q + p_{\bar q})^2 \approx 2 q \cdot p_{\bar q}  \gg \lambdaqcd^2$.
The typical momentum carried by $\bar q$ inside $\overline B$ is of
order $\lambdaqcd\,v$, where $v=p/m_b$ is the $\overline B$  velocity. 
Thus we impose a constraint on the invariant mass of $\overline B$ and $q$
\bea
\label{y2:constraint}
     \overline{y} &\gg &  {\lambdaqcd \, m_b \over M_Z^2} \,.
\eea 
Note this constraint is stronger than $y,\,\overline{y} \gg \lambdaqcd^2/M_Z^2$ in
the $b\overline{b}g$ case.

While these three jets are widely separated in angle, the HQR cross sections
in Eq.~(\ref{ld1s0}) and (\ref{ld3s1}) are  suppressed
by a factor of $\alpha_s(M_Z)\lambdaqcd m_b/M_Z^2\sim 10^{-5}$ relative to $b \overline{b}
g$. 
This should be regarded as a rather conservative estimate, 
since the HQR cross sections in Eq.~(\ref{ld1s0}) and (\ref{ld3s1})
have an extra $1/\overline{y}$ enhancement relative to the $b\bar{b}g$ case.
It is important to investigate the discovery potential 
for this anomalous 3-jet event quantitatively.

Perturbative QCD implies that  
the gluon jet has some characteristic differences from the light quark jet, e.g., 
with higher multiplicity, broader angular distribution  
and softer fragmentation function~\cite{Ellis:qj}.
But it is impossible to distinguish them  
experimentally
on an event-by-event basis due to large fluctuation, 
and only possible on a statistical basis. 
OPAL collaboration has carried out a comparison study at the $Z$ resonance
for $b$, $uds$ jets and $g$ jets~\cite{Alexander:1995bk}.
They selected about $3,000$  symmetric 3-jet events, in which 
the most energetic jet with a mean energy 42.8 GeV
is tagged to contain a $b$, the other two low energy jets 
carry equal  energies about  24.4 GeV.
This corresponds to an event topology where 
the angle between the highest energy $b$-jet and each of
the low energy jets is roughly $150^\circ$.
These samples were assumed to all be the $b \bar{b} g$ events,
but now it is clear there must be a small fraction of anomalous
3-jet events misidentified as $b\bar{b}g$.

Since only the $b$ quark was tagged experimentally, 
we need sum over all the $\overline B$ flavors 
emerging from the recombination.
We include all  the ground-state $\overline B$ mesons, but neglect 
those orbitally excited mesons and $b$ baryons. 
Assuming $SU(3)$ flavor symmetry,
we estimate the ratio of the yield for HQR 3-jet events
to that for the $b \overline{b} g $  in such a topology:
% \displaystyle 
\bea
\label{bqbar:highest}
{ \sum_{\overline B} d {\Gamma}[Z^0 \rightarrow \overline{B}+\overline{b}+q]  \over
d {\Gamma}[Z^0 \rightarrow b+\overline{b}+g]}
&\approx &\, 1.6\times 10^{-3} 
\qquad \qquad a) \,,
\nn \\
&\approx &\, 1.0\times 10^{-2}
\qquad \qquad b) \,.
\nn
\label{b:highest}
\eea
The input parameters are $\sin^2 \theta_w=0.23$,
$\alpha_s(M_Z)=0.12$, $M_Z=91.2$ GeV and $m_b=4.5$ GeV.
We have considered two possibilities for the HQR contribution.
In $a)$, the most energetic jet contains $\overline{B}$ emerging from the recombination;
in $b)$, the highest energy jet is initiated by the recoiling $\bar b$ jet. 
The constraint in Eq.~(\ref{y2:constraint}) is satisfied in both cases.
Because the $\overline B$ and $q$ have smaller invariant mass
(small $\overline{y}$) in the latter case,
the 3-jet production rate is larger than that in the former case.
Including both possibilities  $a)$ and $b)$, 
we estimate there are
roughly $35$ anomalous 3-jet events out of 3,000 samples,
seemingly not to be statistically significant.
We hope that prospective Giga-Z experiment with a much larger
number of $Z^0$ samples will confirm their existences.

Though the HQR cross section is highly suppressed for 3-jet,
its magnitude becomes much larger when $\overline{B}$ and $q$ lie in the fragmentation
region,
i.e., with a small invariant mass. 
Therefore, the dominant contribution to the cross section comes from a back-to-back 
two $b$ jet configuration.
This motivates us to examine if this HQR process also contributes to the 
$b$ fragmentation function.

Fragmentation functions can be expressed as universal 
nonperturbative QCD matrix elements~\cite{Collins:1981uw}.
However, there is a dramatic difference between 
light hadron and heavy quarkonium.
While the fragmentation function for $q$, $g$ turning into $\pi$, $K$ 
can only be extracted from the data,
that for a heavy quark into heavy quarkonium, such as
$D_{\overline{b}\rightarrow B_c}$, can be calculated in perturbative
QCD~\cite{Braaten:1993jn}.

One may expect the fragmentation of heavy quark
into heavy-light hadron interpolates between these two cases, {\it un}calculable versus
{\it fully} calculable 
in perturbation theory.
Indeed, we will show that it is {\it partially} calculable in perturbative QCD.
Some recent work on heavy quark fragmentation function 
using Heavy Quark Effective Theory
can be found in Ref.~\cite{Jaffe:1993ie}.

HQR contribution to the fragmentation function can be obtained
by integrating the double differential cross sections
(\ref{ld1s0}) and (\ref{ld3s1}) over $\overline{y}$ in the limit  
$\gamma\rightarrow 0$.
Naively, one may think the HQR process doesn't contribute to the fragmentation function,
due to the overall factor $\gamma$  in Eq.~(\ref{ld1s0}) and (\ref{ld3s1}).
However, this explicit suppression can be compensated by 
the enhancement of cross sections in the fragmentation region.
To see this,
note the lower integration boundary of $\overline{y}$ 
can be  well
approximated by  $\gamma\,y/(1-y)$.
Carrying out the integration, changing variable 
$y$ to $z$, we have
\bea
\label{frag_my1}
D_{b \rightarrow \overline{B}}^{\rm HQR}(z) &=&
 {32 \,\rho_1^b \,\alpha_s^2(m_b)\over 81}
\,{z\,(2-2\,z+\,z^2) \over (1-z)^2}\,,
\\
D_{b \rightarrow \overline{B}^*}^{\rm HQR}(z) &=&
{32 \, \rho_1^b \, \alpha_s^2(m_b)\over 81}
\,{z\,(2-2\,z+3\,z^2) \over (1-z)^2}\,.
\label{frag_my3}
\eea
We have also approximated the upper boundary of $\overline{y}$  to be $\infty$, 
introducing an error of order $m_b^2/M_Z^2$.
The renormalization scale here must be taken of order $m_b$, 
instead of $M_Z$.
Note the absolute normalization of the above functions is fixed.
Also note these expressions don't depend on the
couplings of $Zb\bar{b}$ at all, 
enforced by that fragmentation functions must be process-independent. 
In fact, 
due to the 
``similarity" between HQR and heavy quarkonium production~\cite{Braaten:2001bf}, 
these functions can be obtained
by a shortcut.
Simply replacing $f_{B_c}/m_c$ in the $D_{\bar b\rightarrow B_c(B_c^*)}$ 
derived in \cite{Braaten:1993jn} by 
$4\sqrt{\rho_1}$ to absorb the singularity from vanishing $m_c$, then 
one can safely take the limit $m_c\rightarrow 0$ to 
recover Eq.~(\ref{frag_my1}) and (\ref{frag_my3}).

The HQR fragmentation functions for both 
$B$ and $B^*$ are not away from zero
until $z$ becomes large, 
and finally badly diverges as $z\rightarrow 1$.
As a result, the total HQR cross section for inclusive $B$ production
is {\it linearly} divergent. 
The breakdown of perturbative calculation as  $z$ approaches 1
is not hard to imagine. 
When $z\sim 1$, $b$  only experiences a slight deceleration, 
and can readily  pick up a $\bar q$ from the vacuum to hadronize.
This should be the dominant contribution to the fragmentation process
in the endpoint region, 
which is a genuine nonperturbative problem, 
unlikely to be tackled in any perturbative framework. 
This is in a sharp contrast to the $B_c$ production,
where the charm is too heavy to be excited from the vacuum, therefore
free from this complication.

We can derive the range of $z$  in which the
perturbative calculation is valid.
Note that $y$  must be large enough so that
for this given $y$, the minimum of $\overline{y} \approx \gamma\,y$  
satisfies the constraint (\ref{y2:constraint}). 
Thus we must have
\bea
\label{frag:valid}
     y &=& 1-z \gg   {\Lambda_{\rm QCD}  \over m_b} \,.  
\eea 
Note this constraint naturally arises from
the fundamental parameters in QCD
and doesn't 
depend on the external parameter $M_Z$.

The $z$ distributions in Eq.~(\ref{frag_my1}) and (\ref{frag_my3}) 
are much harder than the widely used phenomenological 
Peterson fragmentation function~\cite{Peterson:1983ak}.
This may suggest the soft ``vacuum" effects 
are still important,
even in the perturbatively valid $z$ region. 
However, as was advocated in Ref.~\cite{Cacciari:2002pa},
a heavy quark fragmentation function should be decomposed into a 
perturbative part and a nonperturbative part.
While the perturbative component systematically takes into account the
resummation effects associated with $b$ production, 
the nonperturbative one characterizes the universal hadronization
of $b$ to $\overline B$.
It should be kept in mind, however, the nonperturbative
part alone is not a physical observable and 
depends on how one defines the perturbative part.
We will neglect this subtlety.
A recent extraction of the nonperturbative part of fragmentation function 
shows also a harder spectrum than 
Peterson parameterization~\cite{Ben-Haim:2003yu}.

The nonperturbative physics starts to take over when $z\sim 1-\lambdaqcd/m_b$. 
We may
imagine this is a ``matching" point where perturbative contributions
roughly match the nonperturbative contributions. It is also natural to assume
this is where the peak is located~\cite{Jaffe:1993ie}.
Therefore,  the magnitude of $D^{\rm HQR}_{b\rightarrow \overline{B}}(1-\lambdaqcd/m_b)$ 
should be of the same order as the maximum of the ``true" fragmentation function. 
Based on this assumption, a direct observation from 
Eq.~(\ref{frag_my1}) and (\ref{frag_my3}) is that the maximum of 
fragmentation function 
for $\overline{B}^*$ is 3 times larger than that for $\overline B$, 
as expected from spin counting. 
From Eq.~(\ref{frag_my3}), we can estimate
the peak of the fragmentation function 
of $b$ to $B^{*-}$
is located at $z=0.93$, with a height roughly 
\beq
\label{peak:Bstar}
    {96 \over 81} \,\rho_1^b \,\alpha_s^2(m_b)\, 
     \left({m_b \over \lambdaqcd}\right )^2\, \approx  \, 1.5\,,     
\eeq
where we take the running coupling $\alpha_s(m_b)=0.24$, 
$\lambdaqcd=300$ MeV.
This may be an overestimate for $\overline{B}_s^*$, since
the $SU(3)$ flavor symmetry is broken.
If we approximate the ``true" $B^{*-}$ fragmentation function
by Peterson function $D(z;\epsilon_b)$  
with $\epsilon_b=0.006$~\cite{Biebel:2001ka}, and take the fragmentation 
probability $f_{b\rightarrow B^{*-}}\approx 0.3$~\cite{Ackerstaff:1996gz},
the ``true" peak is also around $z \approx 0.93$ with a 
height about 1.7, in good agreement with  
the naive estimate in Eq.~(\ref{peak:Bstar}). 

Since $\rho_1^b \propto \lambdaqcd/m_b$, 
we expect the maximum of the fragmentation function scales
as $\alpha_s^2(m_b)\,m_b /\lambdaqcd$. 
Therefore, the ratio of the maximum of $D_{b\rightarrow B^{*-}}$
to that of $D_{c\rightarrow D^{*0}}$ reads
(Note $B^{*-}$ and $D^{*0}$ are in the same heavy meson multiplet) 
\beq
\label{scaling}
  { \alpha_s^2(m_b)\, m_b \over  \alpha_s^2(m_c)\, m_c} \, \approx \, 1.3\,,   
\eeq
where $m_c=1.5$ GeV and $\alpha_s(m_c)=0.37$  are chosen.
The dependence of the ratio on $\lambdaqcd$ is merely from the
running strong coupling.
With $\epsilon_c=0.05$~\cite{Biebel:2001ka} and 
$f_{c\rightarrow D^{*0}}\approx 0.24$~\cite{Ackerstaff:1996gz},
the Peterson function $D_{c\rightarrow D^{*0}}$ 
peaks around $z=0.8$,  with a height about 0.64.
The ratio of the maxima of the ``true" fragmentation functions
is $1.7/0.64\approx  2.6$, twice as large as
expected from the scaling law in Eq.~(\ref{scaling}).  
Regarding the potentially 
large $1/m_c$ corrections, we are 
content with this crude agreement. 

In summary, we have investigated heavy meson production 
at $e^+e^-$ collider from a new hadronization mechanism.
This mechanism predicts a new type of
3-jet event, power-suppressed relative to the ordinary $b\overline{b}g$ jet.
Even though its discovery is hampered by the
experimental difficulty to distinguish the
quark and gluon jets,
it is possible that future Giga-Z experiment 
will provide
overwhelming evidences for this mechanism.

It is shown that HQR also has 
a
contribution not
suppressed by inverse powers of $M_Z$, 
but only by $\lambdaqcd/m_b$.
We identify this with
a contribution to heavy quark 
fragmentation function.
We have calculated it in the zeroth order of 
heavy quark expansion.
It will be very interesting to  incorporate the $1/m_b$ 
corrections.
The perturbative calculation breaks down 
when the energy fraction  $z$
is close to $1-\lambdaqcd/m_b$. 
Finally we argue that the maximum of the 
fragmentation function for S-wave heavy meson
may respect a simple scaling law. 

One important issue we haven't addressed 
so far
is how to extract the {\it finite} power corrections 
from the power
divergent total HQR cross section. 
To do so, we must identify and then subtract off all the
leading power (fragmentation) contributions in the infrared region. 
This
is a nontrivial task and may need invoke
a more general factorization theorem~\cite{twist3}.

I thank J.~Collins for clarifying some pictures on power corrections.
This work is supported by the 
National Science Foundation under Grant No. PHY-0100677.


\begin{references}

%\cite{Collins:gx}
\bibitem{Collins:gx}
J.~C.~Collins, D.~E.~Soper and G.~Sterman, in
%``Factorization Of Hard Processes In QCD,'' 
{\it Perturbative \ Quantum \  Chromodynamics} (Ed. A.~H.~Muller),
World Scientific (1989).
%%CITATION = 00319,5,1;%%

%\cite{Politzer:1980me}
\bibitem{Politzer:1980me}
H.~D.~Politzer,
%``Power Corrections At Short Distances,''
Nucl.\ Phys.\ B {\bf 172}, 349 (1980);
%%CITATION = NUPHA,B172,349;%%
%
%\cite{Jaffe:1982pm}
%\bibitem{Jaffe:1982pm}
R.~L.~Jaffe and M.~Soldate,
%``Twist Four In Electroproduction: Canonical Operators And Coefficient Functions,''
Phys.\ Rev.\ D {\bf 26}, 49 (1982);
%%CITATION = PHRVA,D26,49;%%
%
%\cite{Ellis:1982wd}
%\bibitem{Ellis:1982wd}
R.~K.~Ellis, W.~Furmanski and R.~Petronzio,
%``Power Corrections To The Parton Model In QCD,''
Nucl.\ Phys.\ B {\bf 207}, 1 (1982).
%%CITATION = NUPHA,B207,1;%%

%\cite{Aitala:1996hf}
\bibitem{Aitala:1996hf}
E.~M.~Aitala {\it et al.}  [E791 Collaboration],
%``Asymmetries between the production of D+ and D- mesons from 500 GeV/c pi- nucleon interactions as a function of xF and pt**2,''
Phys.\ Lett.\ B {\bf 371}, 157 (1996);
M.~I.~Adamovich {\it et al.}  [WA89 Collaboration],
%``Charge asymmetries for D, D/s and Lambda/c production in Sigma-  nucleus interactions at 340-GeV/c,''
Eur.\ Phys.\ J.\ C {\bf 8} (1999) 593;
%%%
F.~G.~Garcia {\it et al.}  [SELEX Collaboration],
%``Hadronic production of Lambda/c from 600-GeV/c pi-, Sigma- and p  beams,''
Phys.\ Lett.\ B {\bf 528}, 49 (2002).



%\cite{Bodwin:1994jh}
\bibitem{Bodwin:1994jh}
G.~T.~Bodwin, E.~Braaten and G.~P.~Lepage,
%``Rigorous QCD analysis of inclusive annihilation and production of heavy quarkonium,''
Phys.\ Rev.\ D {\bf 51}, 1125 (1995)
[{\it Erratum-ibid.} \ D {\bf 55}, 5853 (1997)].
%[hep-ph/9407339].
%%CITATION = HEP-PH 9407339;%%


%\cite{Braaten:2001bf}
\bibitem{Braaten:2001bf}
E.~Braaten, Y.~Jia and T.~Mehen,
%``B production asymmetries in perturbative QCD,''
Phys.\ Rev.\ D {\bf 66}, 034003 (2002);
{\it ibid.} \ D {\bf 66}, 014003 (2002);
%``The leading particle effect from heavy-quark recombination,''
Phys.\ Rev.\ Lett.\  {\bf 89}, 122002 (2002);
%\cite{Braaten:2003vy}
E.~Braaten {\it et al.},
%``Lambda/c+ / Lambda/c- asymmetry in hadroproduction from heavy-quark  recombination,''
hep-ph/0304280.


%%CITATION = HEP-PH 0205149;%%
%\cite{Berger:1980qg}
\bibitem{Berger:1980qg}
E.~L.~Berger, T.~Gottschalk and D.~W.~Sivers,
%``A Higher Twist Term In Inclusive Pion Production At Large Transverse Momentum,''
Phys.\ Rev.\ D {\bf 23}, 99 (1981).
%%CITATION = PHRVA,D23,99;%%

%\cite{Hagiwara:fs}
\bibitem{Hagiwara:fs}
K.~Hagiwara {\it et al.}  [Particle Data Group Collaboration],
%``Review Of Particle Physics,''
Phys.\ Rev.\ D {\bf 66}, 010001 (2002).
%%CITATION = PHRVA,D66,010001;%%

%\cite{Chang:2003ag}
\bibitem{Chang:2003ag}
C.~H.~Chang, J.~P.~Ma and Z.~G.~Si,
%``A QCD analysis of quark recombination for leading particle effect,''
hep-ph/0301253.


%\cite{Isgur:vq}
\bibitem{Isgur:vq}
N.~Isgur and M.~B.~Wise,
%``Weak Decays Of Heavy Mesons In The Static Quark Approximation,''
Phys.\ Lett.\ B {\bf 232}, 113 (1989);
%``Weak Transition Form-Factors Between Heavy Mesons,''
{\it ibid.}\ B {\bf 237}, 527 (1990).


%\cite{Ellis:qj}
\bibitem{Ellis:qj}
R.~K.~Ellis, W.~J.~Stirling and B.~R.~Webber,
{\it QCD and Collider Physics}, Cambridge (1996).
%Cambridge Monogr.\ Part.\ Phys.\ Nucl.\ Phys.\ Cosmol.\  {\bf 8}, 1 (1996).
%%CITATION = 00315,8,1;%%


%\cite{Alexander:1995bk}
\bibitem{Alexander:1995bk}
G.~Alexander {\it et al.}  [OPAL Collaboration],
%``A Comparison of b and (u d s) quark jets to gluon jets,''
Z.\ Phys.\ C {\bf 69}, 543 (1996).
%%CITATION = ZEPYA,C69,543;%%

%\cite{Collins:1981uw}
\bibitem{Collins:1981uw}
J.~C.~Collins and D.~E.~Soper,
%``Parton Distribution And Decay Functions,''
Nucl.\ Phys.\ B {\bf 194}, 445 (1982).
%%CITATION = NUPHA,B194,445;%%



%\cite{Braaten:1993jn}
\bibitem{Braaten:1993jn}
%\cite{Chang:bb}
%\bibitem{Chang:bb}
C.~H.~Chang and Y.~Q.~Chen,
Phys.\ Rev.\ D {\bf 46}, 3845 (1992)
[{\it Erratum-ibid.} \ D {\bf 50}, 6013 (1994)];
%
E.~Braaten, K.~M.~Cheung and T.~C.~Yuan,
%``Perturbative QCD fragmentation functions for B(c) and B(c)* production,''
Phys.\ Rev.\ D {\bf 48}, 5049 (1993).


%\cite{Jaffe:1993ie}
\bibitem{Jaffe:1993ie}
R.~L.~Jaffe and L.~Randall,
%``Heavy quark fragmentation into heavy mesons,''
Nucl.\ Phys.\ B {\bf 412}, 79 (1994);
%%%%%%%%%%%%
E.~Braaten {\it et al.},
%``Perturbative QCD fragmentation functions as a model for heavy quark fragmentation,''
Phys.\ Rev.\ D {\bf 51}, 4819 (1995);
%%%%%%%%%%%%
G.~T.~Bodwin and B.~W.~Harris,
%``Compatibility of various approaches to heavy-quark fragmentation,''
{\it ibid.} \ D {\bf 63}, 077503 (2001).


%\cite{Peterson:1983ak}
\bibitem{Peterson:1983ak}
C.~Peterson {\it et al.},
%``Scaling Violations In Inclusive E+ E- Annihilation Spectra,''
Phys.\ Rev.\ D {\bf 27}, 105 (1983).

%\cite{Cacciari:2002pa}
\bibitem{Cacciari:2002pa}
M.~Cacciari and P.~Nason,
%``Is there a significant excess in bottom hadroproduction at the  Tevatron?,''
Phys.\ Rev.\ Lett.\  {\bf 89}, 122003 (2002);
P.~Nason and C.~Oleari,
%``A fixed-order calculation of the heavy quark fragmentation function in  e+ e- collisions,''
Phys.\ Lett.\ B {\bf 447}, 327 (1999);
M.~Cacciari,hep-ph/0205326.

%\cite{Ben-Haim:2003yu}
\bibitem{Ben-Haim:2003yu}
E.~Ben-Haim {\it et al.},
%``Extraction of the x-dependence of the non-perturbative QCD b-quark  fragmentation distribution component,''
hep-ph/0302157.


%\cite{Biebel:2001ka}
\bibitem{Biebel:2001ka}
O.~Biebel, P.~Nason and B.~R.~Webber,
%``Jet fragmentation in e+ e- annihilation,''
hep-ph/0109282.


%\cite{Ackerstaff:1996gz}
\bibitem{Ackerstaff:1996gz}
K.~Ackerstaff {\it et al.}  [OPAL Collaboration],
%``B* production in Z0 decays,''
Z.\ Phys.\ C {\bf 74}, 413 (1997);
%\cite{Gladilin:1999pj}
%\bibitem{Gladilin:1999pj}
L.~Gladilin,
%``Charm hadron production fractions,''
hep-ex/9912064.
%%CITATION = HEP-EX 9912064;%%

\bibitem{twist3}
J.~C.~Collins, private communication.

\end{references}
\end{document}